# Controlling valley-specific light emission from monolayer MoS$_2$ with achiral dielectric metasurfaces

Yin Liu[a,§,#,*], Sze Cheung Lau[b,§], Wen-Hui Sophia Cheng[c], Amalya Johnson[a], Qitong Li[a], Emma Simmerman[b], Ouri Karni[b,e], Jack Hu[a], Fang Liu[d], Mark L. Brongersma[a], Tony F. Heinz[b,e] and Jennifer A. Dionne[a,*]

[a]Department of Materials Science and Engineering, Stanford University, Stanford, CA 94305, USA

[b]Department of Applied Physics, Stanford University, Stanford, CA 94305, USA

[c]Department of Materials Science and Engineering, National Cheng Kung University, Tainan 701, Taiwan

[d]Department of Chemistry, Stanford University, Stanford, CA 94305, USA

[e]SLAC National Accelerator Laboratory, 2575 Sand Hill Road Menlo Park, CA 94025, USA

*Emails: yliu292@ncsu.edu and jdionne@stanford.edu

[§]Y.L. and S.C.L. contributed equally to this paper

# Present address:

Department of Materials Science and Engineering, North Carolina State University, Raleigh, NC 27606, USA

**Abstract:**

Excitons in two-dimensional transition metal dichalcogenides have a valley degree of freedom that can be optically accessed and manipulated for quantum information processing. Here, we integrate $MoS_2$ with achiral silicon disk array metasurfaces to enhance and control valley-specific absorption and emission. Through the coupling to the metasurface Mie modes, the intensity and lifetime of the emission of neutral excitons, trions and defect bound excitons can be enhanced, while the spectral shape can be modified. Additionally, we demonstrate the symmetric enhancement of the degree-of-polarization (DOP) of neutral exciton and trions via valley-resolved PL measurements, and find that the DOP can be as high as 24% for exciton emission and 34% for trion emission at 100K. These results can be understood by analyzing the near-field impact of metasurface resonators on both the chiral absorption of $MoS_2$ emitters as well as the enhanced emission from the Purcell effect. Combining Si-compatible photonic design with large-scale (mm-scale) 2D materials integration, our work makes an important step towards on-chip valleytronic applications approaching room-temperature operation.

**Main Text:**

In two-dimensional transition metal dichalcogenides (TMDCs), carriers and excitons at K and K' valleys can be characterized by a valley pseudospin. This new degree of freedom holds promise for quantum information processing and storage[1–4]. Valley-polarized excitonic states in these materials couple selectively to chiral photon states through valley circular dichroism. Readout and control of the valley states can therefore be achieved using circularly polarized light (CPL) with different helicity. The exciton-photon interaction also enables the use of the valley states for chiral quantum light emitting devices[5]. Valley-selective photoluminescence (PL) can be quantified by the degree of polarization (DOP), defined as $[I(\sigma^+) - I(\sigma^-)]/[I(\sigma^+) + I(\sigma^-)]$, where $I(\sigma^{+/-})$ is intensity of photoluminescence (PL) with right or left circular polarization. The magnitude of DOP is limited by both valley-specific absorption, as well as decoherence processes, such as intervalley scattering, hindering valleytronics applications at room temperature.

To enhance valleytronic properties, the integration of TMDCs with various plasmonic nanostructures and metasurfaces has been reported.[6,7] For example, plasmonic nanowire waveguides and metasurfaces have been used to increase the exciton emission intensity, and to drive the emission from excitons at different valleys in different directions; this spatial separation of emission is desirable for the readout of valley information[7–11]. In parallel, chiral metasurfaces and nanostructures have been used to enhance valley polarization of PL so that significant valley polarization can be retained at room temperature[13–17]. This enhancement results from the amplified chiral field of the metasurface, which boosts radiative decay of chiral photon states through the chiral Purcell effect[13,18]. However, valleytronic application of these systems are currently limited by three factors: 1) The valley polarization is enhanced when the helicity of the excitation light matches the chirality of the metasurface, but suppressed if helicity of the excitation is switched. 2) The resonance of chiral structures can notably change the polarization of the excitation in the near field, generating excitons at both valleys even when the incident light is purely circularly polarized. 3)

Chiral metasurfaces can significantly modify the polarization of exciton emission, complicating readout of valley information via DOP measurements.

In this work, we integrate exfoliated $MoS_2$ monolayers with an achiral dielectric metasurfaces composed of Si disk arrays (Fig. 1a). The achiral disks are designed to host both electric and magnetic dipoles, which when overlapped, preserve the handedness of the incident illumination. By varying the disk diameter, we find that the intensity, lifetime, spectral shape of the emission of neutral excitons, trions and defect bound excitonic states in $MoS_2$ can be enhanced and modified through coupling to metasurface Mie modes. We also show that our achiral metasurfaces exhibit equal enhancement of valley-polarized PL under CPL excitation with different helicities, in stark contrast with previous studies [13,15]. For optimized metasurfaces, we find that the excitonic lifetime is shortened by 40% and the photoluminescence (PL) intensity of $MoS_2$ is increased more than 30-fold at room temperature. At 100K, the DOP is also as high as 24% for excitons and 34% for trions. When combined with full-field simulations, our DOP trends reveal contributions from both the control of spin-polarization of the excitation field as well as enhanced radiative decay (eg, a Purcell enhancement) by the metasurfaces (Fig.1b).

Si nanodisk arrays were fabricated from a single-crystal silicon layer on a sapphire substrate by e-beam lithography and dry etching (see Methods). Three sets of metasurfaces were fabricated on three separate silicon-on-sapphire chips with different thickness: two with a disk height of $h$ = 100 nm and one with $h$ = 96 nm. Each set contains 12-15 metasurfaces, where each metasurface has a different disk diameter $d$, ranging from 210 to 285 nm. All metasurfaces have a fixed pitch of 350 nm and an array area of 100 um by 100 um. The metasurfaces with $h$ =100nm were used for non-polarization resolved measurements while the metasurfaces with $h$ = 96 nm were used for polarization resolved measurements. Figure 1c shows top-down SEM images of three nanodisk arrays with varying diameters. Cross-sectional SEM imaging (Fig. 1c) confirms the disk geometry and the presence of sharp side walls.

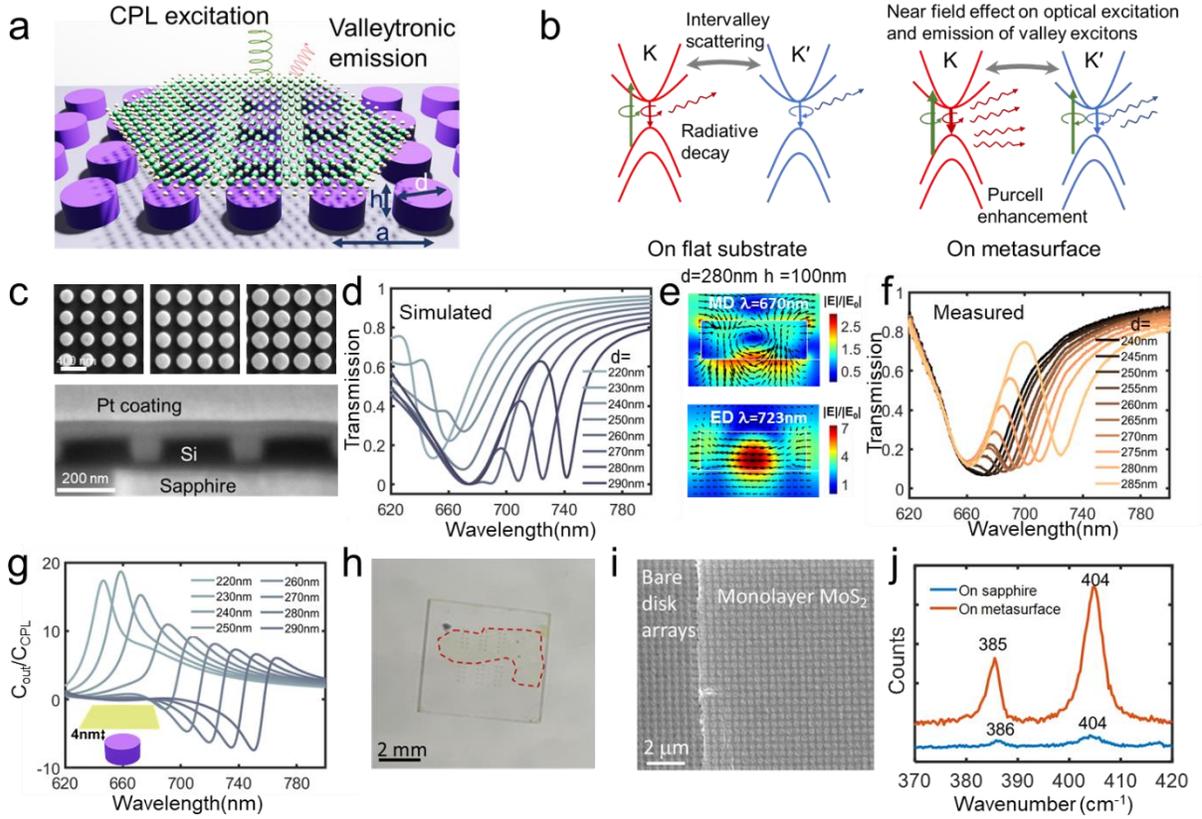

Figure 1 **Coupling monolayer MoS$_2$ with achiral dielectric metasurfaces for controlling valleytronic emission of TMDCs.** (a) Schematic showing the interface of a single layer TMDC with a Si metasurface. (b)Schematic showing the excitation and decay of valley excitons with significant intervalley scattering (Left) and near-field effect of achiral metasurface on the generation and emission of valley excitons (Right). (c) Top-down (upper) and cross-sectional SEM images (lower) of Si disk arrays. Cross sections of disks were created by focused ion beam milling and a Pt layer was deposited for protection only in the SEM process. (d)Simulated transmission spectra of Si dielectric arrays with h = 100 nm and radius 210–290 nm, where the dips result from electric dipole and magnetic dipole resonances. (e) Electric field intensity and vector map of the metasurface ED and MD mode. (f) Measured transmission spectra of Si disk arrays with varying diameters. (g) Spectra of enhanced optical chirality ($C/C_{CPL}$) for varying disk diameters. The inset highlights the plane above the disk arrays where C is calculated. (h,i) Optical photo(h) and SEM image(i) showing a large-scale single layer MoS$_2$ transferred on metasurfaces. The monolayer MoS$_2$ flake is highlighted by red dashed line (j) Raman spectra of single-layer MoS$_2$ on metasurface and sapphire respectively.

We simulated the transmission spectra of the metasurfaces with $h$ = 100 nm and $d$ = 220-290 nm using full-field, finite-difference time-domain (FDTD) simulations (Fig. 1d). Two dips can be identified in the calculated transmission spectra, corresponding to electric dipole (ED) and magnetic dipole (MD) Mie resonances, respectively. We confirm the dipolar nature of the two modes by plotting the magnitude and direction of the electric field, shown in Fig. 1e. As we change the aspect ratio of the disks by varying the diameter, the resonant frequencies shift. The measured transmission of our fabricated metasurfaces (Fig. 1f) exhibits excellent agreement with the simulated results. As the disk diameter decreases from 290 nm, both resonances shift to shorter wavelengths. The electric dipole mode shifts faster than the magnetic dipolar mode, and the two dips start to overlap. At a diameter of 240 nm, the two modes overlap at a wavelength of 660 nm, close to A-exciton resonance of $MoS_2$ at room temperature. The overlapping of modes also leads to increased transmission, indicating a first Kerker-like condition is approached [19–21].

The ED and MD resonances modulate the circular polarization state of the near field with respect to the circular polarization state of incident light. This effect is directly related to the local density of chirality (C), defined as $C = -\frac{\omega}{2c^2} Im(E^* \cdot H)$. Here, E and H represent the complex electric and magnetic fields, Im denotes the imaginary part, and ω and c are the angular frequency and speed of light, respectively. Figure 1g shows the calculated maximum of C outside the silicon disks as we vary the radius and incident wavelength. Importantly, as shown previously by our group, the overlap of ED and MD modes in Si disks enables a uniform-sign enhancement in C.[19,20,22] This preservation of photon spin enables improved valley-specific absorption, while the enhanced fields also boost the Purcell effect, discussed in detail below.

$MoS_2$ monolayers were exfoliated from bulk single crystals using the gold tape method, yielding high-quality single-crystal monolayers with a lateral size of a few millimeters squared[23]. The monolayers were then transferred onto the metasurface (see Methods for details). Such large-scale exfoliated samples enable the coverage of many sets of metasurfaces on a single chip by the same monolayer flake and therefore allow direct comparison of the results from different metasurfaces. Figure 1h includes a photograph of a single $MoS_2$ flake spanning 27 metasurfaces. Although some discontinuities are present, most metasurfaces are uniformly covered by the $MoS_2$ monolayer after the transfer. No significant strain is expected in the layer, except at the edge of the metasurface, due to the relatively high filling ratio of the patterns; this result is confirmed by SEM imaging (Fig. 1i), which reveals a flat and smooth layer with little corrugation on top of the disks. Figure 1j displays representative Raman spectra collected from the MoS$_2$ monolayer on the metasurface with $d$ = 245 nm and on the sapphire substrate near the metasurface. The Raman peaks at 386 cm$^{-1}$ and 404 cm$^{-1}$ in the spectrum collected from MoS$_2$ on sapphire are attributed to E$^1_{2g}$ and A$_{1g}$ phonon modes, respectively, and the 18 cm$^{-1}$ frequency separation between these two modes confirms the single layer nature of the specimen[24]. Interestingly, the Raman intensity is also enhanced by ~10 times on the metasurface, as a combined result of the metasurface enhancement on both the pump field and the Stokes field.

We note that the optical interaction of the MoS$_2$ monolayer with the metasurface is in the weak coupling regime. We simulated the transmission spectra of the monolayer on the metasurface using the dielectric function of MoS$_2$ (S.I. Fig. 1a). [25] The transmission spectra is dominated by the overlapped ED and MD modes of metasurface in the presence of MoS$_2$, whereas the transmission decreases slightly and the transmission dip exhibits a blueshift of 2 nm (S.I. Fig. 1b). There is no signature of anti-crossing behavior or Rabi splitting that would be expected if the exciton and Mie modes coupled strongly to form mixed polaritonic states.[26,27]

The interaction of the MoS$_2$ with metasurfaces was examined using photoluminescence spectroscopy. Figure 2a shows the PL spectra of MoS$_2$ on silicon arrays with varying diameters d under 532 nm excitation at room temperature. With d increasing from 240 nm to 285 nm, the PL intensity and spectral shape changes significantly with an increasing number (1-3) of emission peaks, as a result of the coupling of excitons with radiative metasurface modes. To clarify the correlation, we deconvolute the PL spectra using three Lorentzian functions representing the A-exciton mode, ED mode, and MD mode (S.I. Fig 2). For metasurfaces with modes overlapping or closely spaced in the spectra, one or two Lorentzian functions were used for the fit, assuming that the metasurface modes dominate the emission peak in contrast to the exciton. The mode wavelengths extracted from fitting the PL spectra are shown in Figure 2b, which shows a clear correspondence with the wavelength of dips resulting from ED and MD modes in transmission spectra (Fig. 1f) of the bare metasurfaces without $MoS_2$. (Note that the metasurface with d =280 nm was not covered by $MoS_2$ after the transfer and only data from the transmission spectrum is shown.)

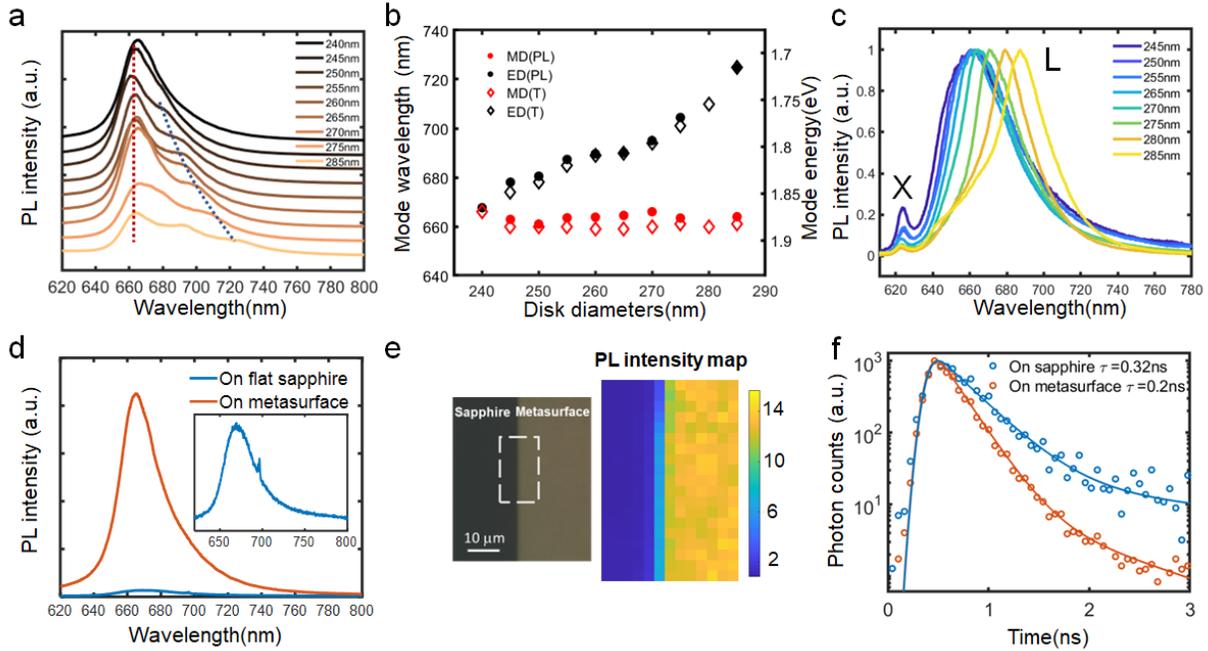

Figure 2 **Modulation of PL spectral shapes and intensity of excitonic emission by metasurfaces** (a) PL of $MoS_2$ on metasurfaces with varying diameters at room temperature. (b) Fitted energy of radiative metasurface modes versus disk diameters. (c) Normalized PL spectra of $MoS_2$ on metasurfaces with varying disk diameters at 5K. (d) Photoluminescence spectra of $MoS_2$ on metasurface with d = 245 nm and flat sapphire. Inset: PL of $MoS_2$ on sapphire (e) Optical image showing $MoS_2$ covering both metasurface and sapphire (left) and the corresponding PL intensity map(right). The map area is marked by the dashed box in the optical image. (f) Time resolved PL of an encapsulated $MoS_2$ on the metasurface and on flat sapphire.

We also measured the PL spectra at temperature 5K (Fig. 2c). The PL peak of neutral A excitons shifts from 665 nm at room temperature to 624 nm at 5K. In addition, a broad defect emission peak (denoted as the L band), resulting from defect bound excitonic states, is present and dominates the emission spectra at 5K. In some reports at cryogenic temperatures, the strain generated by depositing TMDCs on nanopillars could have localized excitons, giving rise to single photon emission with narrow emission linewidth [28,29].

Such emission was not observed in our samples, consistent with the low strain in MoS$_2$ deposited on high-filling-ratio metasurfaces. The defect bound states are coupled to the Mie modes which enhances their emission intensities. The emission peak redshifts from 665 to 696 nm with increasing disk diameters, in correspondence to the redshift of the ED and MD modes in the metasurfaces. These results demonstrate how Mie modes can spectrally control excitonic emission from 2D materials.

At the optimized disk diameter (d=245nm), the PL intensity of MoS$_2$ is enhanced by 30 times on the metasurface compared to a flat sapphire substrate at room temperature (Fig.2d). The intensity map spanning both the metasurface and the sapphire substrate (Fig.2e) shows the clear enhancement and only slight variation in intensity within the metasurface, suggesting that the monolayer and metasurface have uniform optical properties. The enhancement in PL is attributed to enhancement in both laser excitation and exciton emission by the optical near field around the metasurfaces. We performed FDTD simulation and calculated the absorption of MoS$_2$ on metasurfaces and on sapphire substrate respectively and the result suggests ~3 times enhancement of absorption on the metasurface under 532 nm excitation (S.I. Fig 3). The result is also consistent with the measured ~9 times enhancement in the Raman intensity on the metasurface (Fig. 1j) since $I_{abs} \propto |E_{exc}|^2$ whereas $I_{Raman} \propto |E_{exc}|^4$ (given that both the incoming and scattered photon energies are on the far tail of the metasurface resonance). Our analysis therefore suggests the remaining 10 times enhancement is from the Purcell effect on the radiative decay of excitons.

Time-resolved PL measurements were performed to verify the Purcell effect using a 485 nm pulsed laser with a 60 ps pulse width and a repetition frequency of 80 MHz at room temperature (see Methods in SI). No PL was detected from gold tape exfoliated MoS$_2$ monolayers on metasurfaces by limited excitation power (<10 uW ) of the pulsed laser. We do not expect exciton-exciton annihilation in this situation since no PL was detected with CW laser of the same power at room temperature either. Therefore, we encapsulated a scotch tape exfoliated MoS$_2$ monolayer with a-few-nanometer-thick hBN on both sides and transferred it to a metasurface with d = 245 nm for the time-resolved PL measurement. The hBN encapsulated sample exhibits measurable PL on both flat substrate and the meta surface. Figure 2f shows the normalized PL intensity dynamics of the neutral excitons of MoS$_2$ monolayers on sapphire and on the metasurface respectively at room temperature. The time-resolved PL can be fitted using a Gaussian response function convoluted with a biexponential decay function of $Ae^{-t/\tau 1} + Be^{-t/\tau 2}$. For each curve, two distinct time constants in the biexponential function are extracted, i.e., a fast decay ($\tau_1$) of 0.2 ns (metasurface) & 0.32 ns (sapphire) and slow decay ($\tau_2$) of 0.94 ns (metasurface) & 3 ns (sapphire). The fast decay is attributed to the recombination of neutral free excitons, whereas the slow decay is attributed to the recombination of defect trapped or phonon scattered excitons[29]. The decay time of neutral free excitons is determined by both radiative and nonradiative recombination processes as $\frac{1}{\tau_1} = \frac{1}{\tau_r} + \frac{1}{\tau_{nr}}$. With hBN encapsulation, we expect the nonradiative decay time to be very similar on both the metasurface and on sapphire. Therefore, the reduction in overall decay time from 0.32 ns on sapphire to 0.2 nm on metasurface is mainly due to the reduction of radiative decay time from the Purcell effect of the metasurface. Note that the recombination is still dominated by the nonradiative process, especially at room temperature. Although the non-radiative lifetime varies from sample to sample, we may estimate its order of magnitude by $\tau_{nr}$ = 0.1 $\tau_r$ as reported by other experimental studies[31,32]. With such assumptions, we estimate that the radiative recombination time $\tau_r$ on metasurface is 0.13 times $\tau_r$ on sapphire. This number agrees with the ~10-times Purcell factor deduced from simulation and measured Raman intensity mentioned earlier.

Next, we study how our metasurfaces tailor emission polarization from valley excitons. In our measurements, 633 nm and 610 nm circularly polarized laser excitations are used to create excitons at specific valleys. We then analyze the circular polarization state of emitted photons using the degree of polarization (DOP) defined earlier. A schematic of the setup for valley resolved measurement is shown in S.I. Figure 4. Figure 3 (and S.I. figure 5) show the CP-resolved PL spectra and DOP spectra of MoS$_2$ on metasurfaces with varying disk diameters and h=96 nm using 610 nm circularly polarized excitation. Two peaks at 630 nm and 642 nm can be clearly resolved in the spectra, which are attributed to emission from neutral excitons and trions respectively. Under excitation of CPL with different handedness, the spectral shapes and intensities are almost the same except for the flip of polarization sign, giving rise to symmetric DOPs for two excitation helicities. This demonstrates the power of our achiral metasurfaces to achieve symmetric control of the DOP and hence emission from each valley, in contrast to the chiral structures of previous studies[13,15].

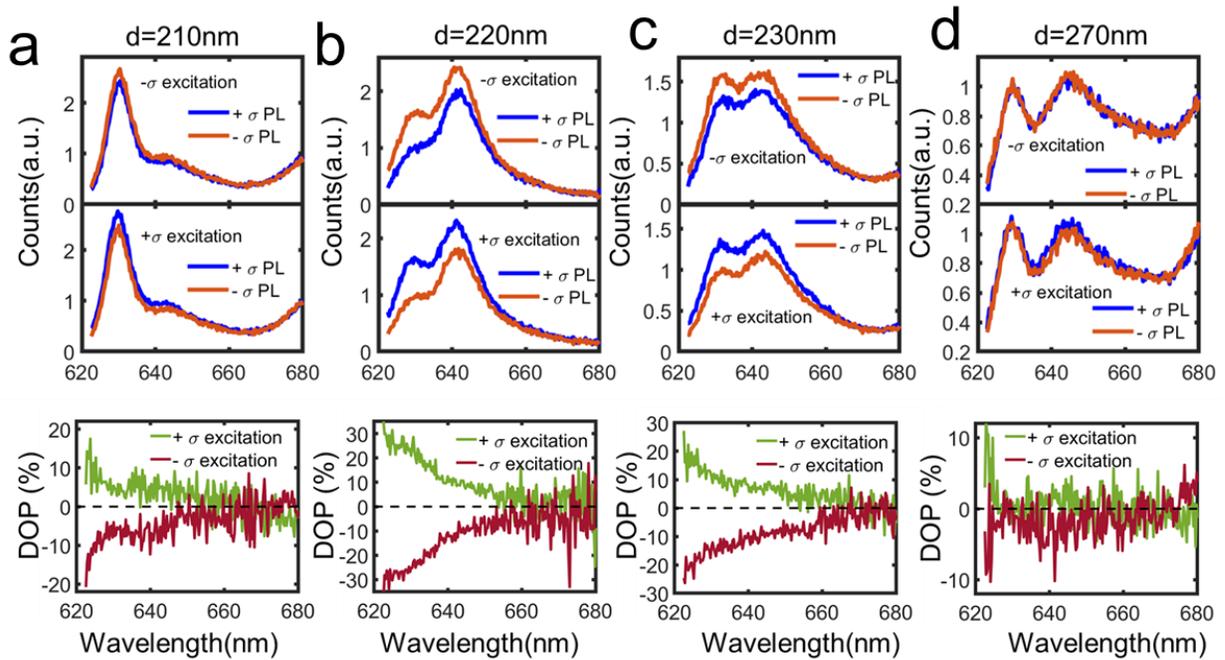

Figure 3 **Valley-resolved PL of excitons and trions under 610 nm excitation** (a-d) Detected circular polarized PL (upper panel) and degree of polarization (lower panel) upon left-handed and right-handed circularly polarized laser excitation respectively for $MoS_2$ on a metasurface with d = 210, 220 230 and 270nm.

We plot the DOPs measured at the emission wavelength of excitons (630 nm) and trion emissions (642 nm) for different metasurfaces, showing a clear dependence of DOPs on the disk diameter (Fig.4). The DOPs of excitons and trions show a maximum of 24% at d =220 and 9 % at d= 230 nm respectively; both drop to zero with d deviating from these values. In addition to the 610 nm excitation, we also measured CPL resolved PL using 633 nm excitation. In this case, the PL of excitons is cut off at 639 nm by a long pass filter and only the DOP of trions at 642 nm was measured (S.I. Fig 6). The resonant coupling of the 633 nm laser with excitons leads to significantly higher PL intensity and larger DOP overall (S.I. Fig 7 and S.I. Fig 8). Figure 4a plots the dependence of the trion DOP at this 633nm excitation wavelength on the disk diameters, showing a maximum value to be 34% at d=220-230 nm. The trend is very similar to the trend obtained from 610 nm excitation, showing maximum DOPs when the MD and ED modes overlap. As a reference, we measured the CPL-resolved PL and DOP of a $MoS_2$ monolayer on a flat, featureless silicon substrate (S.I. Fig.9). The emission is weak, with the PL intensity less than 1% of the PL intensity acquired on our metasurfaces. With 633 nm laser excitation, a DOP of 6% was obtained at 642nm (the trion energy). The magnitude is significantly lower than the maximum 34% DOP obtained from the metasurfaces, verifying the DOP enhancement by the metasurface.

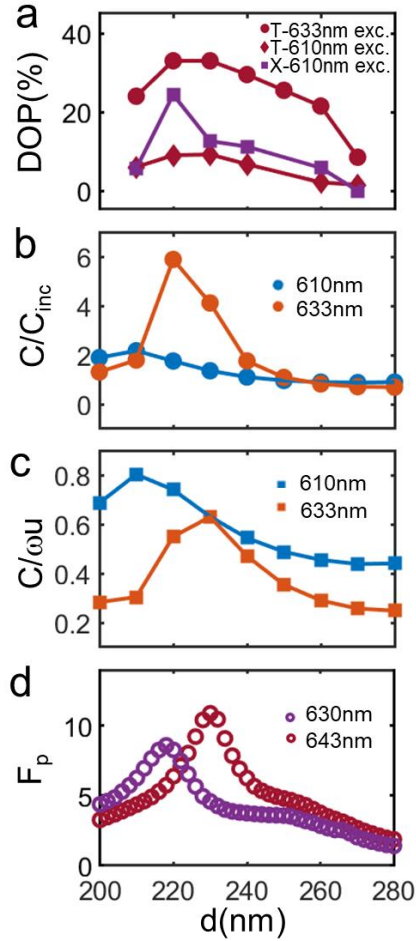

Figure 4 (a)Measured trion and exciton DOP excited with 610 nm and 633 nm lasers. (b)$C/C_{inc}$ at 610 nm and 633 nm as a function of disk diameter d. C is the optical chirality averaged over a unit cell of the metasurfaces (c) $C/\omega u$ as a function of d (d) calculated Purcell enhancement versus d.

We analyze two near-field effects of the Mie modes to understand the disk-diameter-dependent DOP. The first effect is that the metasurface modulates (or enhances) the circular polarization state of the excitation field, potentially boosting valley-specific exciton generation. The second effect is the enhanced radiative decay of excitons due to enhanced optical field at emission wavelength. The measured DOP is determined by $DOP = \frac{P_0}{1 + \tau/\tau_s}$, where $P_0$ is the initial polarization, $\tau$ is the exciton lifetime and $\tau_s$ is the valley depolarization time. $P_0$ represents the polarization modulation by the metasurfaces, expressed by $P_0 = 2\alpha - 1$, where α is the percent of excitons generated at one valley under CPL illumination. When excitons are only excited in one valley, $P_0$ equals one. The exciton lifetime is determined by radiative and nonradiative recombination time, $\frac{1}{\tau} = \frac{1}{\tau_r} + \frac{1}{\tau_{nr}}$, which can be shortened via decreasing the radiative recombination time $\tau_r$ by metasurfaces.

To understand the change of the initial polarization, we calculate the quantity C/ωu that characterizes the circular polarization state of the excitation field, where C is the optical chirality, ω is the angular frequency, and u is the energy density of the electromagnetic field.[33] For an optimal chiral near field that fully preserves the chirality of incident CPL, C/ωu =1. C/ωu thus characterizes the differential excitation of excitons in K and K' valleys. Figure 4b and 4c plot C and C/ωu as a function of disk diameter under 610nm and 633 nm CPL excitation. As seen, this quantity approaches 1 and is maximized between d=210nm, and 230nm, depending on the excitation wavelength. Therefore, the disks are both enhancing the spin-selective absorption (C), and the rate of excitation of valley-specific excitons.

Finally, to illustrate the metasurface's effect on the exciton emission, we calculate the radiative enhancement of $MoS_2$ on metasurfaces using COMSOL. The room-temperature dielectric function of silicon was used in the simulation. This is justified by our measured reflection spectra of bare metasurfaces at room temperature and 100K, showing very little change in the spectra shape governed by Mie resonances of the nanodisks (S.I. Fig 10). To represent the valley excitons, we use a cloud of in-plane chiral dipoles formed by pairs of two orthogonal electric dipoles with a relative phase difference of 90 degrees. 100 chiral dipoles are placed in random orientation and phases on the surface above the nanodisk in a unit cell. The Purcell Factor $F_p$ of the metasurface is calculated by dividing emitted power with the presence of metasurface by the emitted power without metasurface. The dependence of $F_p$ on disk diameters at the peak wavelength for trion (642nm) and exciton (630 nm) is shown in Figure 4d.

Overall, the disk diameter dependence of the measured DOP can be well correlated to the disk diameter dependence of $C/\omega u$ and $F_p$. The disk diameter corresponding to maximums DOP for excitons and trions occur in between the disk diameters corresponding to the maximum values of $C/\omega u$ and $F_p$. The modulation of DOP is therefore attributed to the polarization modulation of the excitation field and the enhancement of radiative recombination on the metasurfaces.

In summary, we have shown the first large-scale integrated $MoS_2$ monolayer – metasurface systems that enable enhanced valleytronic absorption and emission. We show that the emission intensity and spectral shape of $MoS_2$ can be markedly modified via the coupling of neutral excitons, trions or defect bound excitonic states with the metasurfaces. Strong enhancements in the emission intensity and reductions in the exciton lifetime are observed at frequencies where metasurface EDs and MDs are overlapped. We also observed a strong, symmetric enhancement in the DOP. Our results can be accurately attributed to a combination of enhanced spin-selective absorption, valley-specific exciton generation, and Purcell effect. Further improvement to room temperature operation could be achieved by high-qualify-factor metasurfaces. We anticipate such structures will enable enhanced valley-specific absorption, exciton generation-rates, and Purcell enhancements at least two orders of magnitude larger than that the current low-Q metasurfaces of this study[34–37].

**Supporting information**

Methods for metasurface fabrication, sample preparation and optical measurements; FDTD simulation of transmission spectra of metasurfaces with and without monolayer $MoS_2$; fitting of the PL spectra of MoS2 on metasurfaces; Simulated absorption spectra of $MoS_2$ on metasurfaces; non-polarized and circularly polarized PL measurements of $MoS_2$ on metasurfaces and flat silicon at 100K under 610 nm and 633 nm laser excitation; reflection spectra of bare metasurfaces with different disk diameters at room temperature and 100K.

**Author information**

Y. L. and S.C.L. contributed equally to this work.

**Acknowledgements**

The authors acknowledge support from the U.S. Department of Energy, Office of Science, Office of Basic Energy Sciences, under award no. DE-SC0021984, which supported the salaries of Y. L, K. S, M. B. and J. D. In addition, the authors acknowledge support for 2D materials synthesis from the Q-next grant under award no. DE-AC02-76SF00515. T. H. salary was supported by the Moore Foundation grant under award no. 10146. W.-H.C. acknowledges the support from Ministry of Science and Technology, Taiwan (2030 Cross-Generation Young Scholars Program, MOST 110-2628-E-006-007), and Ministry of Education (Yushan Fellow Program), Taiwan, and in part from the Higher Education Sprout Project of the Ministry of Education to the Headquarters of University Advancement at National Cheng Kung University (NCKU). The authors thank Dr. Yongmin Liu at Northeastern University for valuable feedback on the work.

# References

1. Mak, K. F., He, K., Shan, J. & Heinz, T. F. Control of valley polarization in monolayer MoS2 by optical helicity. *Nat. Nanotechnol.* **7**, 494–498 (2012).

2. Zeng, H., Dai, J., Yao, W., Xiao, D. & Cui, X. Valley polarization in MoS2 monolayers by optical pumping. *Nat. Nanotechnol.* **7**, 490–493 (2012).

3. Xu, X., Yao, W., Xiao, D. & Heinz, T. F. Spin and pseudospins in layered transition metal dichalcogenides. *Nat. Phys.* **10**, 343–350 (2014).

4. Wang, G. *et al. Colloquium* : Excitons in atomically thin transition metal dichalcogenides. *Reviews of Modern Physics* **90**,021001(2018)

5. Schaibley, J. R. *et al.* Valleytronics in 2D materials. *Nature Reviews Materials* **1**, 1–15 (2016).

6. J. Jang *et al.* Planar optical cavities hybridized with low-dimensional light-emitting materials, *Adv Mater,* 35, 2203889(2023).

7. Y. Chen *et al.* Multidimensional nanoscopic chiroptics, *Nature Reviews Physics* 4, 113(2022).

8. Gong, S.-H., Alpeggiani, F., Sciacca, B., Garnett, E. C. & Kuipers, L. Nanoscale chiral valley-photon interface through optical spin-orbit coupling. *Science* **359**, 443–447 (2018).

9. Sun, L. *et al.* Separation of valley excitons in a MoS2 monolayer using a subwavelength asymmetric groove array. *Nat. Photonics* **13**, 180–184 (2019).

10. Hu, G. *et al.* Coherent steering of nonlinear chiral valley photons with a synthetic Au–WS2 metasurface. *Nat. Photonics* **13**, 467–472 (2019).

11. Bucher, T. *et al.* Tailoring Photoluminescence from MoS2 Monolayers by Mie-Resonant Metasurfaces. *ACS Photonics* **6**, 1002–1009 (2019).

12. Cihan, A. F., Curto, A. G., Raza, S., Kik, P. G. & Brongersma, M. L. Silicon Mie resonators for highly directional light emission from monolayer MoS2. *Nat. Photonics* **12**, 284–290 (2018).

13. Li, Z. *et al.* Tailoring MoS2 valley-polarized photoluminescence with super chiral near-field. *Adv.*

*Mater.* **30**, e1801908 (2018).

14. Lin, W.-H. *et al.* Electrically tunable and dramatically enhanced valley-polarized emission of monolayer WS2 at room temperature with plasmonic Archimedes spiral nanostructures. *Adv. Mater.* **34**, e2104863 (2022).

15. Wen, T. *et al.* Steering valley-polarized emission of monolayer $MoS_2$ sandwiched in plasmonic antennas. *Science Advances*, **6**, eaao0019 (2020).

16. Wu, Z., Li, J., Zhang, X., Redwing, J. M. & Zheng, Y. Room-temperature active modulation of valley dynamics in a monolayer semiconductor through chiral Purcell effects. *Adv. Mater.* **31**, e1904132 (2019).

17. Guddala, Bushati, Li & Khanikaev. Valley selective optical control of excitons in 2D semiconductors using a chiral metasurface. *Opt. Mater. Express.* **9**, 536 (2019).

18. Yoo, S. & Park, Q.-H. Chiral Light-Matter Interaction in Optical Resonators. *Phys. Rev. Lett.* **114**, 203003 (2015).

19. Solomon, M. L., Hu, J., Lawrence, M., García-Etxarri, A. & Dionne, J. A. Enantiospecific Optical Enhancement of Chiral Sensing and Separation with Dielectric Metasurfaces. *ACS Photonics* **6**, 43–49 (2019).

20. Abendroth, J. M. *et al.* Helicity-preserving metasurfaces for magneto-optical enhancement in ferromagnetic $[Pt/Co]_N$ films. *Adv. Opt. Mater.* **8**, 2001420 (2020).

21. Staude, I. *et al.* Tailoring directional scattering through magnetic and electric resonances in subwavelength silicon nanodisks. *ACS Nano* **7**, 7824–7832 (2013).

22. Solomon, M. L., Abendroth, J. M., Poulikakos, L. V., Hu, J. & Dionne, J. A. Fluorescence-Detected Circular Dichroism of a Chiral Molecular Monolayer with Dielectric Metasurfaces. *J. Am. Chem. Soc.* **142**, 18304–18309 (2020).

23. Liu, F. *et al.* Disassembling 2D van der Waals crystals into macroscopic monolayers and reassembling into artificial lattices. *Science* **367**, 903–906 (2020).

24. Li, H. *et al.* From bulk to monolayer MoS2: Evolution of Raman scattering. *Adv. Funct. Mater.* **22**,

1385–1390 (2012).

25. Li, Y. *et al.* Measurement of the optical dielectric function of monolayer transition-metal dichalcogenides:MoS2,MoSe2,WS2, andWSe2. *Phys. Rev. B Condens. Matter Mater. Phys.* **90**, 205422 (2014).

26. Liu, X. *et al.* Strong light–matter coupling in two-dimensional atomic crystals. *Nat. Photonics* **9**, 30–34 (2014).

27. Chen, Y.-J., Cain, J. D., Stanev, T. K., Dravid, V. P. & Stern, N. P. Valley-polarized exciton–polaritons in a monolayer semiconductor. *Nat. Photonics* **11**, 431–435 (2017).

28. Branny, A., Kumar, S., Proux, R. & Gerardot, B. D. Deterministic strain-induced arrays of quantum emitters in a two-dimensional semiconductor. *Nat. Commun.* **8**, 15053 (2017).

29. Palacios-Berraquero, C. *et al.* Large-scale quantum-emitter arrays in atomically thin semiconductors. *Nat. Commun.* **8**, 15093 (2017).

30. Korn, T., Heydrich, S., Hirmer, M., Schmutzler, J. & Schüller, C. Low-temperature photocarrier dynamics in monolayer MoS2. *Appl. Phys. Lett.* **99**, 102109 (2011).

31. Scuri, G. *et al.* Large Excitonic Reflectivity of Monolayer MoSe2 Encapsulated in Hexagonal Boron Nitride. *Phys. Rev. Lett.* **120**, 037402 (2018).

32. Shi, H. *et al.* Exciton dynamics in suspended monolayer and few-layer $MoS_2$ 2D crystals. *ACS Nano* **7**, 1072–1080 (2013).

33. Hanifeh, M. & Capolino, F. Helicity maximization in a planar array of achiral high-density dielectric nanoparticles, *J. Appl. Phys*, 27, 093104 (2020).

34. Hu, J., Lawrence, M. & Dionne, J. A. High Quality Factor Dielectric Metasurfaces for Ultraviolet Circular Dichroism Spectroscopy. *ACS Photonics* **7**, 36–42 (2020).

35. Lawrence, M. *et al.* High quality factor phase gradient metasurfaces. *Nat. Nanotechnol.* **15**, 956–961 (2020).

36. Lawrence, M. & Dionne, J. A. Nanoscale nonreciprocity via photon-spin-polarized stimulated

Raman scattering. *Nat. Commun.* **10**, 3297 (2019).

37. Schneider, G. F., Calado, V. E., Zandbergen, H., Vandersypen, L. M. K. & Dekker, C. Wedging transfer of nanostructures. *Nano Lett.* **10**, 1912–1916 (2010).

# TOC graphic

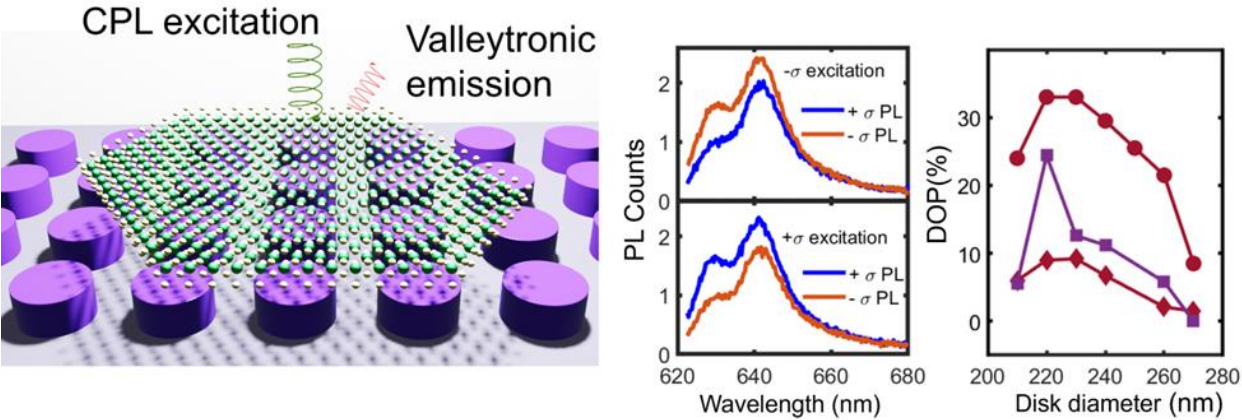